\documentclass[runningheads]{llncs}
\usepackage[misc]{ifsym}
\usepackage{enumitem}
\usepackage{subfigure}
\usepackage{graphicx}
\usepackage{epsfig}
\usepackage{color}
\usepackage{algorithmic}
\usepackage[ruled,vlined]{algorithm2e}
\usepackage{amsmath}
\usepackage{amssymb}
\usepackage{booktabs}
\usepackage{url}
\usepackage{enumitem}
\usepackage[sort]{cite}
\usepackage{bbding}
\usepackage{orcidlink}

\newcommand{\eat}[1]{}

\begin{document}
\title{Online Digital Investigative Journalism using SociaLens}
\author{Hasan M. Jamil\,\orcidlink{0000-0002-3124-3780} \Envelope  \and Sajratul Y. Rubaiat\,\orcidlink{0009-0001-0367-644X} 
}
\authorrunning{Jamil and Rubaiat}
%
\institute{Department of Computer Science, University of Idaho, Moscow, ID 83844, USA\\
\email{jamil@uidaho.edu, ruba3062@vandals.uidaho.edu}}

\titlerunning{SociaLens}

\maketitle

\begin{abstract}
Media companies witnessed a significant transformation with the rise of the internet, bigdata, machine learning (ML) and AI. Recent emergence of large language models (LLM) have added another aspect to this transformation. Researchers believe that with the help of these technologies, investigative digital journalism will enter a new era. Using a smart set of data gathering and analysis tools, journalists will be able to create data driven contents and insights in unprecedented ways. In this paper, we introduce a versatile and autonomous investigative journalism tool, called {\em SociaLens}, for identifying and extracting query specific data from online sources, responding to probing queries and drawing conclusions entailed by large volumes of data using  ML analytics fully autonomously. We envision its use in investigative journalism, law enforcement and social policy planning. The proposed system capitalizes on the integration of ML technology with LLMs and advanced bigdata search techniques. We illustrate the functionality of SociaLens using a focused case study on rape incidents in a developing country and demonstrate that journalists can gain nuanced insights without requiring coding expertise they might lack. SociaLens is designed as a ChatBot that is capable of contextual conversation, find and collect data relevant to queries, initiate ML tasks to respond to queries, generate textual and visual reports, all fully autonomously within the ChatBot environment.
\end{abstract}

\section{Introduction}

The ever growing digital news data enabled the creation of the term ``data journalism" \cite{McGregorH14}. The abundance of journalistic data presents new opportunities. However, it also creates technical hurdles, which have given rise to the idea of computational journalism \cite{CohenLYY11}. The promises of data or computational journalism led researchers develop NewsCollab \cite{RolimAC023} and Crosstown Foundry \cite{NoceraCTKKS21} to support archived news access and use, and also venture into automated news generation \cite{LeppanenMGT17} and news mashups \cite{EkaputraDKNTTA17}. Struggles emerged to identify misinformation \cite{FernandezA18} and assign trust in news \cite{OpdahlTDMSTTT23}, all apparently aided and enabled by the profusion of data, internet and social networks.

Investigative journalism plays a critical role in our social life by shining a light on the good, bad and the ugly. The recent proliferation of misinformation and fake news in digital media make it extremely difficult to separate truth from falsehood, and even harder for investigative journalists to sift through mountainous volumes of data to find the truth. They are now suddenly required to deal with data collection \cite{WiedemannYB18}, data management \cite{AnadiotisBBCGHH21}, data wrangling \cite{KasicaBM21}, and their eventual analysis \cite{FernandesMC23}. While the digital profusion of journalistic data is promising for investigative journalism, it is still challenging for parts of the world where human suffering and oppression are higher due to structural, social and political states \cite{HaqueYASAH20}, especially to combat corruption \cite{JainSCSY2018}, abuse and discrimination \cite{WangAFCC23}, violence against women \cite{CorreaF21} hunger, free speech \cite{De-ArteagaB2019}, and other social concerns. Journalists need more easily accessible sophisticated investigative tools with higher degrees of usability.

Journalists today need a computational platform that provides access to relevant information -- e.g., archived news articles, blogs, databases, records, digital libraries, etc., use data gathering and wrangling tools, analytics, data visualization, and the ability to hypothesize about events and predict possible events. Such a platform will empower them to report impactful and newsworthy pieces for public consumption. In this paper, we present the contours of an intelligent ChatBot, called Socia{\em Lens}. SociaLens is a machine learning based investigative journalism tool that supports autonomous data gathering, relevant feature identification, model generation, prediction and information visualization. It incorporates situational awareness, and cognition-based forecasting of events using automated machine learning from multi-modal data. A user-friendly graphical interface is used to gather facts from free texts in structured form. Complementary stored data and inference rules are used to support contextual conversational question answering that relies on predictive analysis in real time. We use a social science study on child rape incidents in a developing nation to illustrate the functionalities. We show that contextually structured data can be harvested from online articles toward answering user questions that necessitate analysis of multiple data sets using state of the art machine learning tools. Additionally, it allows situational hypothesizing based on query assumptions.

\section{Related Research}

Investigative journalism is a laborious profession. It requires collecting, collating, organizing, summarizing, linking and analyzing various forms of information and data in the process of seeking answers to interesting questions. Recent explosion of digital media, social networks, micro-blogs, and various forms of archived data is transforming the way journalists seek answers to their questions.

The interest in data journalism has steadily grown for more than a decade, and researchers believe that enough progress has been made for investigative journalism to make a true come back \cite{Witte2018}. However, the resurgence is contingent upon access to clean data \cite{KasicaBM23}, tools for data visualization \cite{BarrosB12}, and analytical tools. As observed in \cite{Philp2022}, collecting, collating, organizing and analyzing data is no child's game \cite{KasicaBM21}. It requires substantial training, technical knowledge and sophistication in data management \cite{AnadiotisBBCGHH21}, informational visualization \cite{BraunischRZ18}, especially when advanced data analysis techniques such as machine learning is needed \cite{FernandesMC23}.

In SociaLens, we incorporate three fundamental autonomous technologies for interactive conversation -- a) data identification, gathering and management, b) machine learning for data analysis and prediction, and c) information visualization. The backbone is built upon the large language model, OpenAI GPT-4o \cite{OpenAI_GPT_4o} API, which, combined with our customized agent structure specifically designed for journalists, serves as the natural language conversational agent. Analysis and visualization drivers are integrated with the conversation agent to complete the autonomous data journalism tool box for investigative journalism. 

\section{Tour of SociaLens}
\label{tour}

Consider a group of socially concerned journalists investigating the state of abuse and exploitation of children in a developing nation such as Bangladesh where the institutions of law enforcement and justice are flailing due to lack of democratic institutions, and its political and economic owes. Their goal is to report how and why child rape is increasing, and what measures are socially necessary to cube the ongoing abuse. Reliable data is scant, record keeping are low priority, and trust in the system is weak. Often the only source of incident information is the online new media and social networks severely oppressed under the weight of the violence and retribution by the political forces. Free speech is under threat of the powerful and the government forces where journalists are constantly watching their backs and double checking what they express in fear of retaliation because abuses are often perpetrated by the people of power and politically connected.

\begin{figure}[ht!]
  \centering
  \includegraphics[width=0.85\textwidth, keepaspectratio]{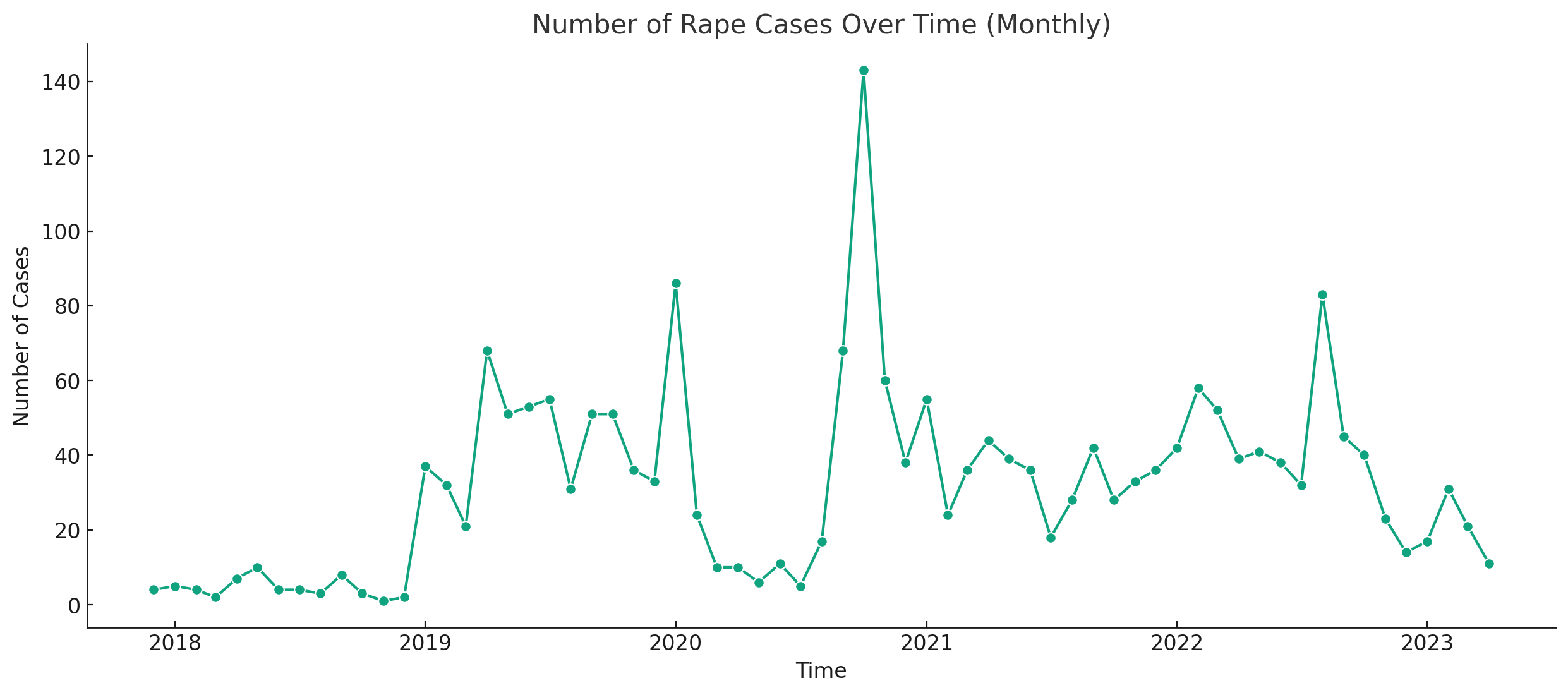}
  \caption{Number of monthly rape incidents between 2018 and 2023.}
  \label{fig:rapecaseyearly}
\end{figure}

SociaLens aims to provide contextual data gathering tools, an ML analysis back-end, generative AI toolbox for query intent recognition and analysis, and a natural language (NL) response generation using an LLM. It uses OpenAI ChatGPT 4o as the conversational agent for query analysis and computational task generation for data gathering, analysis and response construction. Using the array of tools supported, it is able to respond to queries such as
\begin{quote}
    $Q_1$: How often incidents of rape happen in Bangladesh? Could you generate a monthly trend of rape incidents from available reports?
\end{quote}
to generate the following response and a visual plot of rape incidents as shown in Fig \ref{fig:rapecaseyearly}.
\begin{quote}
    Here's the updated line plot showing the monthly trend in the number of rape cases, as reported in the news stories. The trend is similar to what we observed earlier, with peaks around the start of 2020 and mid-2021.
\end{quote}

\begin{figure}[ht!]
  \centering
  \includegraphics[width=0.5\textwidth, keepaspectratio]{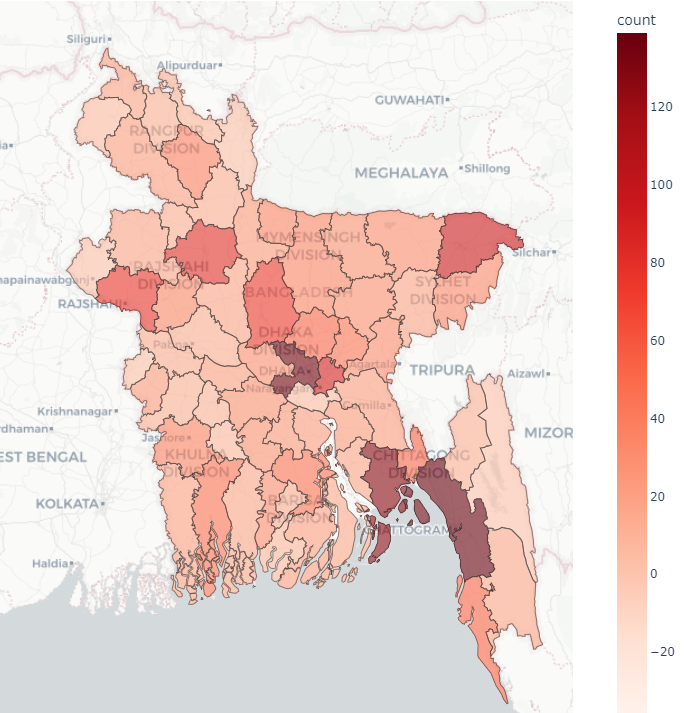}
  \caption{Spatial distribution of reported rape cases in Bangladesh.}
  \label{fig:rapecasehotspotmap}
\end{figure}

\begin{figure}[ht!]
  \centering
  \includegraphics[width=0.85\textwidth,keepaspectratio]{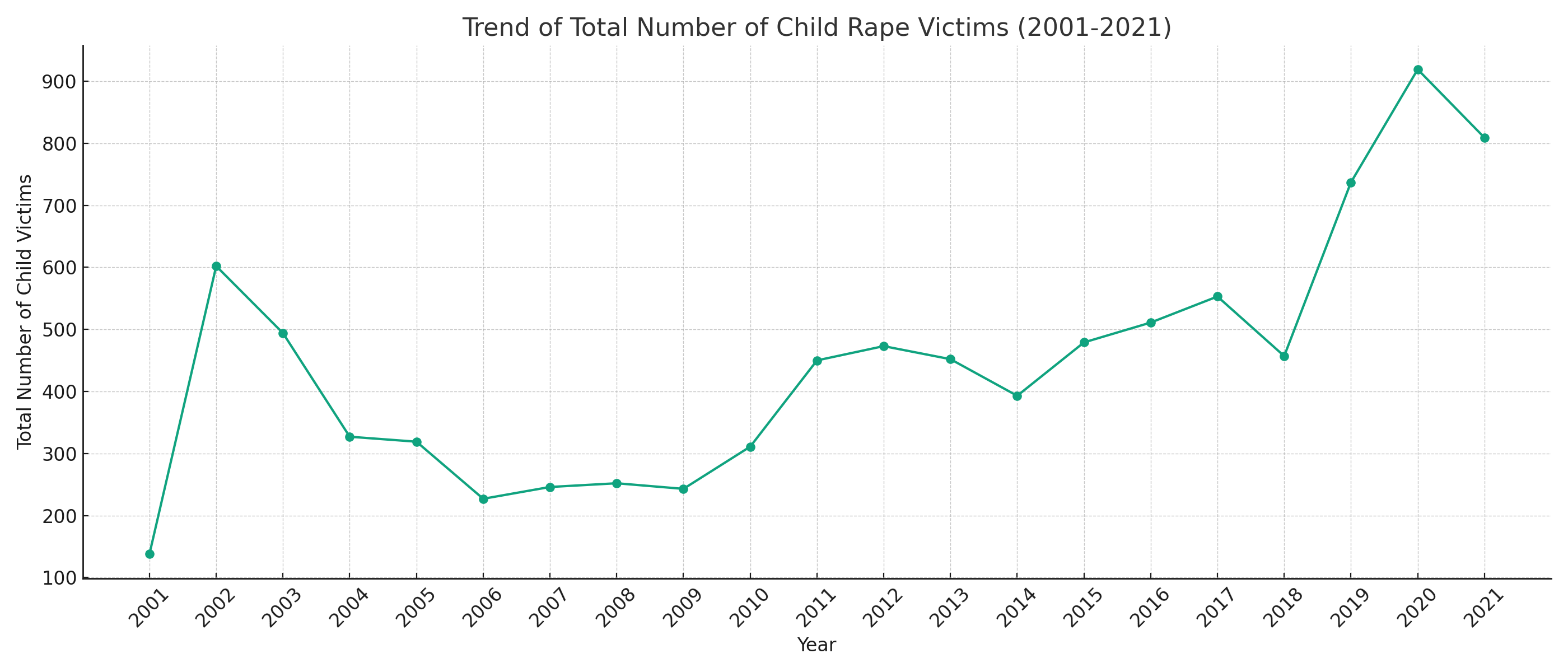}
  \caption{Annual trend of child rape incidents in Bangladesh.}
  \label{fig:childrapecase}
\end{figure}

It is also capable of responding to more involved queries of the form
\begin{quote}
    $Q_2$: Please show the geographic hot spots rape incidents in the country.
\end{quote}
to which it responds by saying
\begin{quote}
    Certainly! I've constructed a district-wise map of rape hot spots (see Fig \ref{fig:rapecasehotspotmap}) for your review. Within this map, each region's color intensity corresponds to the number of reported cases in that district, with darker shades indicating higher incidences.
\end{quote}
Finally, it can also respond to more insightful queries such as
\begin{quote}
    $Q_3$: Is the child rape incidents worsening?
\end{quote}
with the following response.
\begin{quote}
    A concerning uptick in incidents is discernible post-2017, culminating in the distressingly high figures of 2020. The chart in Fig \ref{fig:childrapecase} summarizes the trend from 2001 to 2021. 
\end{quote}
While these queries are conceptually simple and somewhat deceiving of the technical capabilities of SociaLens, the underlying engine is powerful to perform a complex array of data gathering, wrangling and analysis tasks and report generation capabilities to match them. The range of queries actually shows its capability to integrate geospatial data with tabular data, and generate a wide range of multi-modal visual reports. In the next few sections we formally discuss the theoretical basis and technical  organization of SociaLens.

Before we proceed to lay out the formal foundation of SociaLens, it is instructive to note that for the sake of brevity and focus, in the current presentation of SociaLens, we limit our attention to already gathered data sets from national newspapers, and data collected from various non-governmental organizations' (NGO) annual reports working in the country, public surveys, and several other collected and curated data types. We also have the capabilities to connect to fixed set of online news archives such as NewsAPI \cite{newsapi}, NewsData.io \cite{newsdataio}.

We are currently developing an autonomous online data source identification and data extraction system called FAIRyfier. FAIRyfier uses LLM analysis of PubMed abstracts to discover potential relevant biological data sources in the context of a query, followed by a resource capability discovery and access plan generation to be used as a data source in the user query. Discovery of these meta-information in Needle \cite{JamilN23} resource description language is sufficient to generate an extraction plan in structured format using a data integration system called VisFlow \cite{MouJ20}. We omit a discussion on online data discovery and extraction in SociaLens using FAIRyfier and refer the readers to  FAIRyfier for a complete exposition. In the remainder of this paper, we only focus on collected and locally archived data sources without any loss of generality.

\section{Conversational Model of SociaLens}
\label{model}

Formally, the SociaLens ChatBot $\Sigma$ is defined by the tuple of the form \linebreak $\langle \lambda, \Delta, \mu, {\mathcal T},  \phi, \omega, \rho\rangle$, where $\lambda$ is the LLM (ChatGPT 4o), $\Delta$ is the materialized or archived local database of the form $\langle \delta_i, d_i\rangle$, $0\leq i\leq n$, $\mu$ a mapping function of the form $\mu: Q \rightarrow 2^{\mathcal T}$, $\phi$ is a function that associates a set of database resources with a database query in $\Sigma$, and finally, $\rho$ is a NL response generator from a query analysis.

The database $\Delta$ is a set of tables $\delta_i$ with associated NL descriptions $d_i$ in natural English. $d_i$s describe the purpose, capability, columns and the meaning of the columns for each database table $\delta_i$. {\bf Q} is an infinite set of NL queries $q$, where a query $q$ is a set of English sentences. $\mu$ is an LLM agent that maps a query $q$ into a set of tasks $T$ of the form ${\mathcal T}$, i.e., $T\in 2^{\mathcal T}$. Here, in turn, every $t\in T$ is a pair $\langle \chi, \kappa \rangle$, where $\chi$ is a classification of $t$ into ``data" ($\kappa_d$), or ``query" ($\kappa_q)$, and $\kappa$ is a NL description of the data need or query need generated from a $t$ by LLM $\lambda$ as a transformation of $t$. Intuitively, $\mu$ transforms each query into a list $L$ of data needs and query needs expressed in NL. The logical order of the list semantically organizes the computational order and thus the pipeline entailed by a query $q$.

The function $\phi$ identifies every data query $\kappa_d$ into a data resource $\delta_i$ using the techniques of FAIRyfier, and $\omega$ maps every $\kappa_q$ into a list of queries in $\Psi$ where $\Psi= S\cup M$, and $S$ is all possible SQL constructs, and $M$ is the set of all MQL \cite{Jamil24} constructs. MQL is a recently proposed declarative machine learning query language. Technically, $\phi(\kappa_d)=d\in \Delta$, and $\omega(d,\kappa_q)=Q\in 2^\Psi$, i.e., set of structured queries in SQL or MQL. Finally, the response reconstruction function $\rho$ is an LLM agent that constructs a response in NL corresponding to a query $q$. It does so my analyzing the query $q$, its mappings $\phi(\kappa_d)$, $\omega(d,\kappa_q)$ and the execution $\eta(\omega(d,\kappa_q))$.

The semantics assigned to each query $q$ is the function
\begin{equation}
\rho(\eta (\Pi_{i=1}^k(\mu(q)))) \text{~~~~} \forall i, i=1,\ldots, k
\end{equation}
where $\eta$ is the SociaLens execution engine comprised of an MQL engine implemented over an SQL database system, and
\begin{equation}
    \Pi(q_i)=\begin{cases} 
      \emptyset & \text{if } \mu(q)= q_i=\langle \chi_i, \kappa_{d_i}\rangle \\
      \omega(\phi(\kappa_{d_i}), \kappa_{q_i})) & \text{if } t\in\mu(q)= q_i=\langle \chi_i, \kappa_{q_i}\rangle \bigwedge \exists i, \mu(q)= q_i=\langle \chi_, \kappa_{d_i}\rangle
   \end{cases}
\end{equation}
In other words, $\Pi(q)$ is the sequence of SQL or MQL queries executed by $\eta$ and explained by $\rho$ corresponding to query $q$.

\subsection{Machine Learning Query Language}

For the sake of completeness, we briefly introduce the most basic MQL construct called {\tt GENERATE}, and refer the readers to \cite{Jamil24} for a complete exposition to it. Furthermore, since SQL is a known quantity, we omit a discussion on SQL entirely. MQL includes two other statement types  -- {\tt CONSTRUCT} and {\tt INSPECT}. While {\tt CONSTRUCT} statement can be used for model generation for prediction, classification and clustering tasks, {\tt GENERATE} statement is used to execute one of these three ML tasks, either using an already constructed model or using a model generated at run time. The {\tt INSPECT} statement is used only for data wrangling and cleaning tasks.

The general syntax of the {\tt GENERATE} statement is shown below. The {\tt GENERATE} statement stands at the level of SQL's SELECT statement and is the main workhorse of MQL. It operates on tabular data to make predictions, categorize objects and group sets of objects into bins. It has five basic clauses -- an ML class selection (one of PREDICTION, CLASSIFICATION and CLUSTER), optional object labeling, feature selection, a data set, a filter condition over the data set, and an input table of unknown cases (test set).
\begin{verbatim}
   GENERATE [DISPLAY OF] PREDICTION v [OVER O] |
   CLASSIFICATION INTO L1, L2, ..., Lp [OVER O] |
   CLUSTER OF k
   [USING MODEL ModelName | ALGORITHM AlgorithmName]
   [WITH MODEL ACCURACY P] [LABEL B1, B2, ..., Bm]
   [FEATURES A1, A2, ..., An
   FROM r1, r2, ..., rq
   WHERE c]
\end{verbatim}
In the above statement, $r_l$ is a table over the scheme $R_l$, $c$ is a Boolean condition, $A_i \in \cup_l R_l$, $s$ is a table over the scheme $\cup_j B_j \bigcup \cup_i A_i$, k is an integer, and $v \in  \cup_l R_l$, $L_k \in  dom(X)$\footnote{$dom(X)$ is the set of elements in the domain of the column $X$, and $X \in  \cup_i A_i$.}.

$v$ in the PREDICTION clause is the target variable, and $A_i$s are the features. The optional LABEL clause identifies attributes $B_j$ as the object identifiers for all the three ML tasks. The CLASSIFICATION clause classifies each object $\cup_j B_j$ into one of $L_k$ categories. The $k$ in CLUSTER clause is an integer expression that can include SQL aggregate functions over the tables $r_l$. Finally, the optional USING clause is meant to either use an existing model (MODEL option) generated using the CONSTRUCT clause, or a specific ML algorithm (ALGORITHM option) for the generation of the model. As in SQL, WHERE is an optional clause, but unlike SQL, FROM is required. The OVER clause supplies the unknown test dataset over the scheme $A_i \cup B_j$. The ACCURACY option accepts a threshold within the interval (0,1).

\subsection{Implementation of the Execution Engine $\eta$}
\label{imp}

Alg \ref{alg:socialens} captures the spirit of the semantic model of SociaLens. In this algorithm, {\em TaskList} $T$ collects a list of tasks classified by the LLM $\lambda$ using the mapping function $\mu$ either as a data task or as a query task, still in NL form. $\eta$ expects the tasks in $T$ to follow a specific sequence, i.e., a data task followed by a query task which needs the preceding data table, and never in any other order. Therefore, for every $t_i$, $\eta$ either finds the data table using $\phi$, or generates the query expression $e_i\in \Psi$ and queues it in ${\mathcal Q}$. Finally, every query expression in $e\in {\mathcal Q}$ is interpreted by the LLM agent $\rho$ taking into account the original task $t_i$, the generated query expression $e_i\in {\mathcal Q}$ and its executed result $\eta(e_i)$.

\SetKwComment{Comment}{/* }{ */}
\begin{algorithm}
\caption{SociaLens Process $\eta$}\label{alg:socialens}
\KwData{User query $q\in {\bf Q}$;}
\KwResult{Executed results explained by $\rho$;}
\Begin{
    {\em TaskList} $=T\in {\mathcal T}=[t_i=\langle \chi_i, \kappa_{d_i}\rangle]_{i=1}^k \gets \mu(q)$\;
    ${\mathcal Q} \gets []$\; 
    \ForEach{$t_i=\langle \chi_i, \kappa_{d_i}\rangle \in T$}{
        \eIf{$\chi_i=d$}{
            $\phi(\kappa_{d_i})=\delta_i$
            }{
            ${\mathcal Q}\gets \omega(\phi(\kappa_{d_{i-1}}), \kappa_{q_i})$
            }
        }
    $\Sigma \gets \eta({\mathcal Q})$\;
    $Expl \gets \emptyset$\;
    \ForEach{$e_i \in {\mathcal Q}$}{
        $Expl \gets Expl \bigcup \rho(t_i, e_i, \eta(e_i))$
    }
    \Return($\Sigma, Expl$)\;
}
\end{algorithm}

\begin{figure}[ht!]
  \centering
  \includegraphics[width=0.85\textwidth,keepaspectratio]{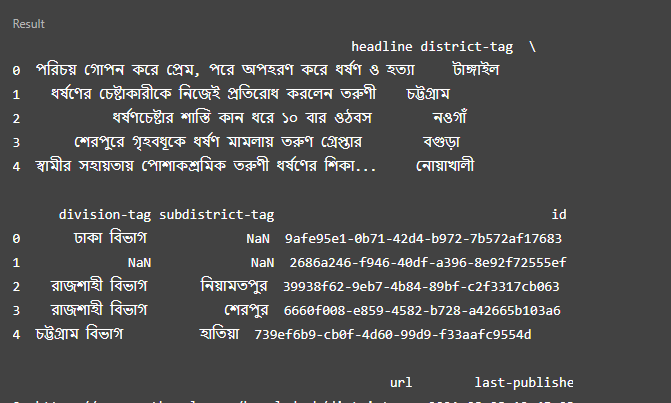}
  \caption{ProtomAlo sample data in native Bangla language.}
  \label{fig:data}
\end{figure}

Let is examine how $\eta$ assigns semantics to query $Q_1$. Once user enters the query text of query $Q_1$ in Sec \ref{tour}
into the chat window, the engine springs into action that is initiated by the system prompt

and proceeds to display the sample data it is using in Fig \ref{fig:data}.
and generates the following explanation.

\begin{figure}[ht!]
  \centering
  \includegraphics[width=.9\textwidth,keepaspectratio]{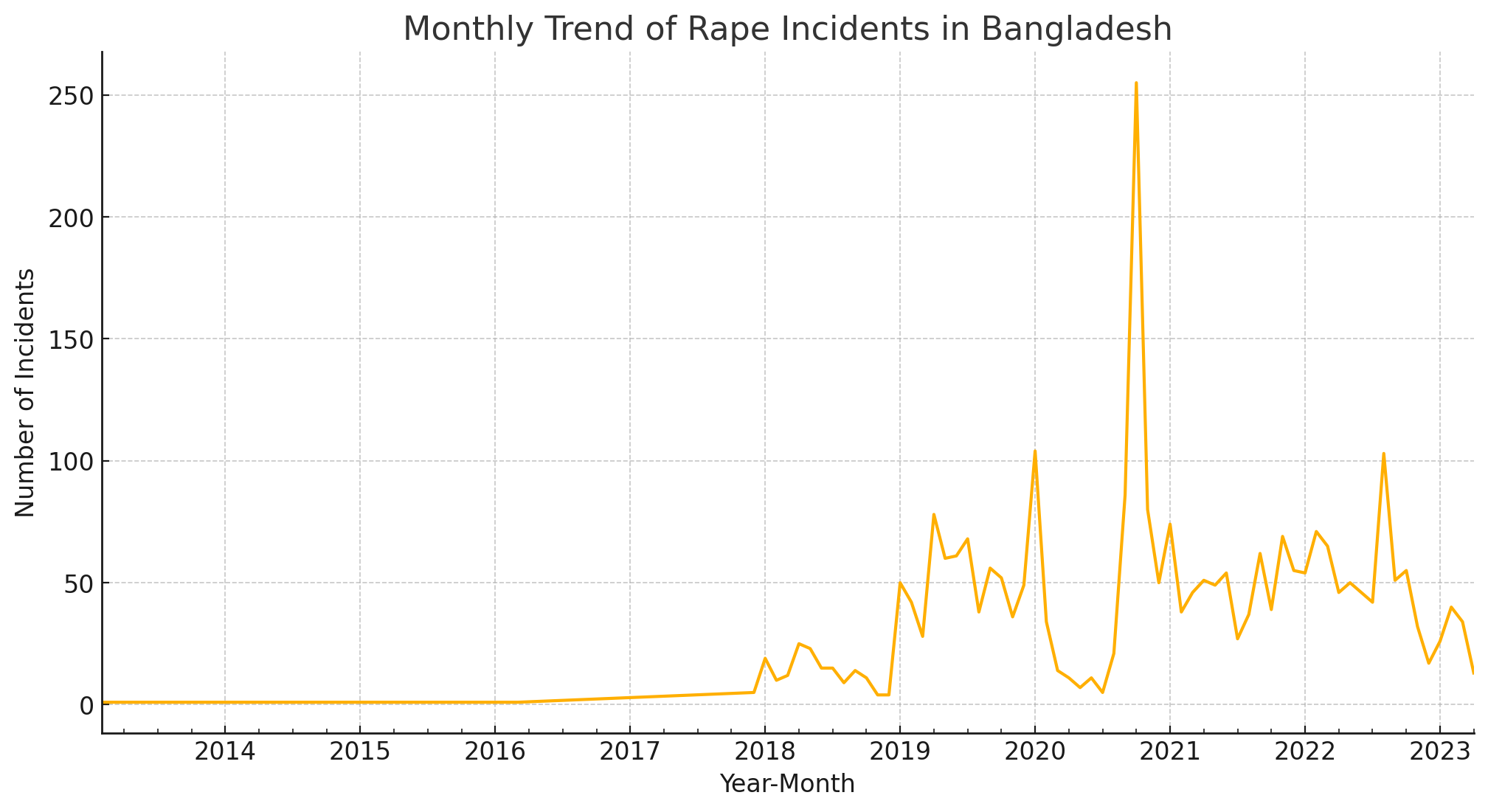}
  \caption{Frequency plot of rape incidents between 2014-2023.}
  \label{fig:graph}
\end{figure}

It then proceeds to execute the following code segment it generated corresponding to the task $\mu(Q_1)=t$.

and generates the trend visualization in Fig \ref{fig:graph} (note the difference in time period with the plot in Fig \ref{fig:rapecaseyearly} due to updates in collected data at two different time of query execution).

It is interesting to note that this particular query does not invite or require the support of SQL or MQL. A simple Python plotting code fragment using matplotlib library is enough. We will revisit the issue of mapping a task to an MQL query in Sec \ref{queryproc}.

\section{Semantics of the Language of $\Sigma$}

We discuss the semantics of a SocialLens query in the context of the three example queries $Q_2$ and $Q_3$ we presented in Sec \ref{tour} to present a conceptual understanding the theoretical model introduced in Sec \ref{model}. As alluded to earlier, for this presentation, we will use publicly available datasets ProthomAlo and NGORep \cite{BDRape2024} and published for the queries $Q_1$ through $Q_4$. While the semantics of SociaLens programs or conversations do not depend on the specific set of tables, these example database and the queries exemplify the semantics assignment process.

\subsection{SociaLens Data Archive}

The pertinent portions of the two data sets \cite{BDRape2024} used in  our queries are summarized below.

{\bf Prothom Alo} is a leading national daily newspaper of Bangladesh. We use reporting of rape incidents in this newspaper as our primary data source. The ProthomAlo dataset is a curated collection of rape incidents reported in Prothom Alo and comprised of 2811 records, each with 8 features -- {\em ID, URL, headline, district-tag, division-tag, subdistrict-tag, last-published-at} and {\em offset}. These features largely includes the title of the news, geographic location of the incident at various granularity, and most recent update timestamp. The record cover the time period from February 22, 2013 until April 10, 2023. 

The second data set, the {\bf NGORep}, is a collection of complementary data retrieved from Kaggle \cite{woman_harassment_dataset_2023}. It consists of 
24,311 records with information on rape attempts from 2014 to 2021, rapes followed by victim's suicide attempts from 2007 through 2021, gang rape incident between 2007 and 2021, incidents of rapes followed by homicides between 2007 and 2021, and total number of rape incidents from 2001 through 2021.


The last, but not the least, we have also linked two online news archives -- NewsAPI \cite{newsapi} and NewsData.io \cite{newsdataio} -- with SociaLens to offer journalists an opportunity to browse current news and emerging trends on incidents of rape and abuse of women using a user friendly interface. These repositories, however, require substantial pre-processing before they can be used in SociaLens' structured archives. Our plan is to leverage the capabilities of FAIRyfier in the near future.

\subsection{Query Processing in SociaLens}
\label{queryproc}

As alluded to in Sec \ref{imp}, we leverage the technique of prompt engineering and retrieval augmented generation (RAG) supported in LLMs, especially in ChatGPT 4o. While a complete discussion of the prompt engineering and RAG technique we follow is beyond the scope of this paper due to space constraints, the discussion in Sec \ref{imp} should have provided some idea how it may have been designed in SociaLens. The discussion in this section, however, depicts that not only we map task lists to SQL or MQL queries (readers may learn about these techniques in \cite{LiHQYLLWQGHZ0LC23} and \cite{Jamil24}, and ommited in this paper), we also map tasks to Python codes, especially for functions not supported in SQL or MQL. For example, the frequency plot in Fig \ref{fig:graph} cannot be generated in SQL or MQL. Furthermore, the geospatial analysis and display shown in Fig \ref{fig:rapecasehotspotmap} cannot also be generated by either of these languages. For these types of analysis and visualization of data, we fall back to Python code fragments. However, for graph generation and visualization using ML analysis in MQL can be easily accomplished. We refer interested readers to \cite{Jamil24} for details, but as an exposition, discuss an intuitive example involving MQL in Sec \ref{mql}.

\subsection{ML Analysis in SociaLens}
\label{mql}

\begin{figure}[ht!]
   \centering
   \includegraphics[keepaspectratio,width=0.65\textwidth]{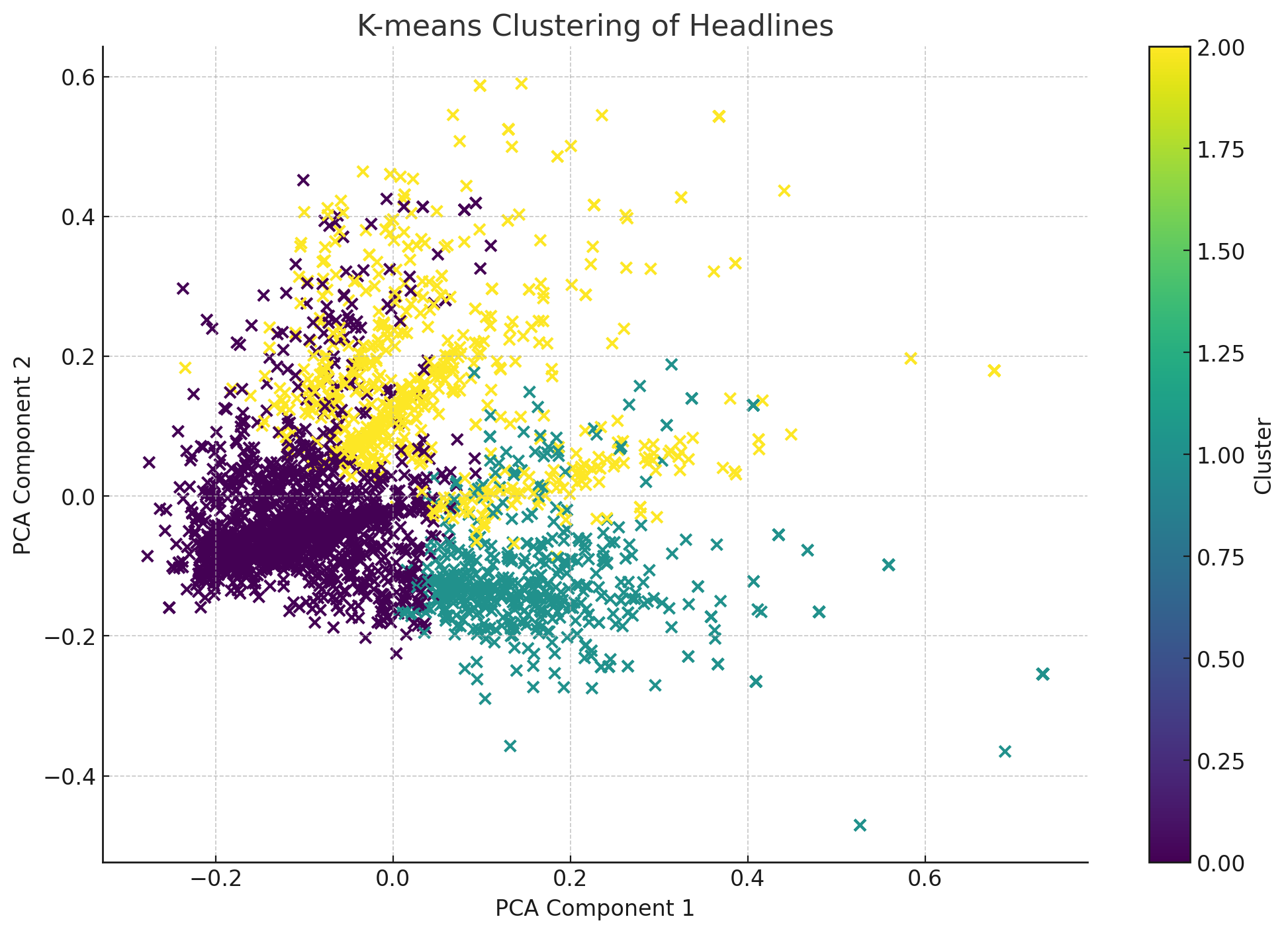}
   \caption{Clustering of Rape Case News Headlines.}
   \label{fig:rape}
\end{figure}

The prothom alo dataset \cite{BDRape2024} includes features such as headline, district name, division name,  publication date, and content. An investigative question could be
\begin{quote}
    {\bf Q4}: What are the top 3 categories of headlines of incidents of rape?
\end{quote}
which will generate and execute the MQL query below using K-means as the default algorithm for clustering (see Fig \ref{fig:rape}). In this example, the default algorithms were chosen since no preference or accuracy threshold was requested.
\begin{verbatim}
   GENERATE DISPLAY OF CLUSTER OF 3 
   ALGORITHM KMeans FEATURES headline 
   FROM ProthomAlo;
\end{verbatim} 

\section{Discussion and Future Research}

SociaLens assembled a number of data journalism tools that interact with the LLM to enable a conversation driven journalistic data analysis and hypothesis testing using machine learning for non computer savvy journalists in the direction of recently introduced ToRA \cite{gou2023tora}. In contrast to ToRA, SociaLens aids smarter conversation and knowledge informed investigation using seamless integration of autonomous machine intelligence, AI and information visualization with LLM. The query processing engine integrates intelligent functions with the LLM agent to recognize ML functions and invoke them as needed. In other words, we are now able to ask ChatGPT questions that require data collection, feature selection, ML analysis, model generation, prediction, and textual summarizing of results, as well as visualize the results. We believe, SociaLens is the first system to do so for investigative journalism.

\subsection{Current Limitations}

The integrated functionalities of MQL though a powerful feature, SociaLens engine cannot yet translate all mapped NL queries into MQL. For example, the query
\begin{quote}
    $Q_5$: Are the rape incidents trending higher in specific regions, and what are the profiles of the victims and the likely perpetrators.
\end{quote}
to highlight what population groups need to be on the alert and watch out for a possible attacker, cannot be directly mapped fully automatically to MQL and supplementary Python codes to answer it. Although additional synthetic data \cite{BDRape-Supple-2024} are available to to compute such queries, the current mapping algorithm $\mu$ needs additional structures, to be able to generate the MQL task lists. Additionally, the MQL language also needs to be strengthened to be able to sync with the conceptual terms such a trend, and profile. Generally, these terms map to a combination of ML tasks that can now be achieved using prompt engineering, most likely on a case by case basis. Developing a general solution will need further research.

Furthermore, the case of profile generation is more complex. It requires isolating the victims that are part of the trend line incidents, and then discovering their general and most common features to build a profile. Some form of feature engineering and analysis could be helpful which MQL currently cannot accommodate directly. Designing a mapping algorithm using the basic MQL features will necessitate complex prompt engineering which remains as our future research goal.

\subsection{Future Plans}

In the second edition of SociaLens, we plan to include support for social leaders and administrators to pose ``what if" type of questions to decide policy discourse against evolving social events, such as increasing incidences of child rape, or violence against women. They should be able to ask questions such as {\em what factors are most likely contributing to increased child rapes}, {\em what law enforcement measure most likely will curb the rapes}, or {\em what is the likelihood of increased policing to reducing the rape incidences}? Such analyses will require intelligent discovery of relevant data, and fitting a predictive curve. The architecture of SociaLens is poised to incorporate these features without much difficulty.

In particular, we believe that contemporary techniques still have limitations, especially to digest and respond using cause-effect relationships and complex societal interactions. This is where SociaLens offers promise and presents an opportunity to fundamentally transform policy-making. We plan to consider the following ideas in our next edition.

\begin{itemize}
\item \textbf{Cause and Effect Analysis:} Governance or what is type of queries of the form ``If we implement stricter regulations on urban migration, how might this impact the frequency of reported rape cases in our city?" should be possible. Such reasoning questions demand the agent to interlace insights across domains, from urban migration patterns to crime rates.

\item \textbf{Scenario Simulations:} A query of the form "If we reduce the population growth rate, how might this affect various socio-economic parameters, including crime rates?" should necessitate SociaLens to extrapolate from current data, offering potential future scenarios spanning areas such as employment, infrastructure demands, and social welfare.

\item \textbf{Recommendation Systems:} Beyond mere analysis, SociaLens could potentially be trained to recommend policy actions. For example, "Given our current socio-economic conditions, suggest policies to alleviate the rape crisis?", SociaLens could suggest actionable strategies rooted in data, historical outcomes, and predictive analysis.

\item \textbf{Cross-Referencing Global Data:} One of the strengths of an LLM-based agent like SociaLens is its capacity to understand and cross-reference large volumes of global information not in SociaLens' archive. Policy makers should be able to improve their policy positions and strategies by leveraging global best practices by querying "How have nations with similar socio-economic conditions addressed their rape crises?" This would enable SociaLens to provide insights that are globally informed yet locally relevant.
\end{itemize}

SociaLens aims to provide journalists with tools to conduct investigations more quickly and efficiently, and to help gain deeper insights. We strive to build an environment that is best suited for investigators, journalists, and policymakers to make well-informed news or decisions.

\section{Conclusion}

SociaLens is a novel embodiment of the integration of LLM and ML technologies in a single system for comprehensive and convenient investigative journalism using a smart ChatBot. Its novelty is in autonomous invocation of ML tasks from the ChatBot using generative AI analysis using ChatGPT 4o. This was possible because of the design of the declarative ML query language MQL. The introduction of MQL and its use in SociaLens leveraged the experience of Text2SQL mapping technology using LLMs \cite{LiHQYLLWQGHZ0LC23}. We have already developed a technique to  map natural language queries into MQL sentences with a high degree of precision. SociaLens thus inherits a robust query processing infrastructure from our previous research on MQL as well as from FAIRyfier and VisFlow and help journalists to decode complex datasets, translating them into compelling and impactful social insights leveraging advanced analytical algorithms with the help of our SociaLens' intuitive  ChatBot interface.

\section*{Acknowledgement}

Research was supported in part by a National Institutes of Health Institutional Development Award (IDeA) \#P20GM103408, a National Science Foundation CSSI grant OAC 2410668, and a US Department of Energy grant DE-0011014.

\bibliographystyle{abbrv}
\bibliography{references,journalism}

\end{document}